\documentclass[lettersize,journal]{IEEEtran}  
\usepackage{amsmath,amsfonts,amssymb,amsthm}
\usepackage{algorithmic}
\usepackage{algorithm}
\usepackage{array}
\usepackage[caption=false,font=normalsize,labelfont=sf,textfont=sf]{subfig}
\usepackage{textcomp}
\usepackage{stfloats}
\usepackage{url}
\usepackage{verbatim}
\usepackage{graphicx}
\usepackage{cite}

\begin{document}

\title{Single vs Multi Vector Predictive Control of Five-phase Drives}

\author{Manuel R. Arahal$^1$, Manuel G. Satué$^1$, Kumars Rouzbehi$^1$, Juana M. Martínez-Heredia$^2$
\thanks{$^1$Department of Systems Engineering and Automation, Universidad de Sevilla, Seville, Spain.}
\thanks{$^2$Department of Electronic Engineering, Universidad de Sevilla, Seville, Spain.}
}

\maketitle

\begin{abstract}
The field of Finite State Model Predictive Control for multiphase drives has produced many contributions. Many variants of FSMPC exist, each aiming at some aspect such as complexity of the cost function, switching frequency, etc. Despite past efforts to compare different techniques, the field is still out of consensus regarding the relative merits of each one. This paper presents a new method to compare FSMPC variants. The method is based on analyzing the modulation, implicit or explicit, used by each variant. In the paper the method is used to compare single-vector state-of-the-art FSMPC with a multi-vector variant designed to cancel $xy$ currents and simplify the cost function. The results show the strengths and weaknesses of each technique. Also, it is found that the trade-offs between figures, previously thought to concern just individual regimes, extend to the whole operating space and also can be pinpoint to each FSMPC variant. Finally, it is shown that the flexibility of the single-vector approach and its better DC-link usage makes it, arguably, superior over the multi-vector variant.
\end{abstract}

\begin{IEEEkeywords}
Multi-phase motor, Power converters, Predictive control, Variable speed drives, Virtual Voltage Vector
\end{IEEEkeywords}

\section{Introduction}
Finite State Model Predictive Control (FSMPC) has been proposed for current control in different electrical systems using a Voltage Source Inverter (VSI) \cite{kennel2000predictive}. The most salient feature of these applications is the elimination of the modulation stage. Pulse Width Modulation (PWM) is no longer necessary as the VSI states are directly set by the FSMPC \cite{lim2022continuous}, \cite{an2024robust} and \cite{liu2024sensorless}.

In Single Vector FSMPC (SV-FSMPC), the selected VSI configuration produces a certain phase voltage known as Voltage Vector (VV) \cite{alharbi2024review}. This corresponds to the basic FSMPC scheme, using just one VV per sampling period \cite{taherzadeh2024six}. Since the control actions are quantized in magnitude and application time, it is obvious that SV-FSMPC achieves some form of modulation. However, the characteristics of said modulation not been studied in a systematic way. This is surprising given the large body of literature that FSMPC has spurred lately \cite{borreggine2019review,xue2023recent}. In fact, some FSMPC variants have been introduced from observations on the sequences of voltage vectors. These papers often aim at reducing the computational load, resorting to reduction of the control set  \cite{mamdouh2022simple} (i.e. the set of allowed  VVs). This avenue is still being pursued despite the recent apparition of computationally light methods \cite{serra2021computationally,arahal2024multi}.

A special case is found when more than one VV is issued per sampling period  \cite{he2024model}. This is the Multi-Vector approach (MV-FSMPC) \cite{li2024novel} in which some form of inter-sample modulation is achieved. Several variants of the multi-vector approach do exist \cite{liu2024three,juan12guiding}, although it can be argued that all of them share some basic principles that are exemplified by the Virtual Voltage Vector (VVV) paradigm also known as Synthetic Vectors \cite{yan2024synthetic}.

Works devoted to modulation are more scarce. For instance, an study of SV-FSMPC is found in \cite{arahal2016harmonic}, where the total harmonic distortion produced by VVs with uniform application times is investigated by exhaustive exploration of VV sequences. The study reveals that, unlike PWM, the modulation of SV-FSMPC features gaps and other undesirable behavior. Of course, for high sampling frequencies (compared with fundamental frequency), these effects are less acute. Unfortunately, the study of \cite{arahal2016harmonic} is prohibitive for high sampling frequencies to fundamental frequency ratios.

\subsection*{Contributions}
This paper presents a new method to compare FSMPC variants. The method is based on analyzing the modulation, implicit or explicit, offered by each variant. The method is used to compare single-vector FSMPC (with and without weighting factors) with a multi-vector designed to cancel $xy$ currents \cite{gonccalves2019finite}.

Please note that the issue of WF tuning has obscured somehow the comparison since multi-vector approaches tend to not use WF. This has been alleviated in recent years with the introduction of novel techniques for WF tuning \cite{yao2023weighting,liu2021neural}. In this sense the paper adds to the on-going discussion about WF and their role in the trade-offs between figures of merit. In particular, it extends previous results obtained for single operating points. In doing so it provides a global assessment of different techniques.

The rest of the paper is organized as follows. The next section presents the elements of the modulation analysis used to compare the single-vector and multi-vector variants of FSMPC for a five-phase Induction Machine (IM). Section 3 presents the results of the modulation analysis, followed by Section 4 where a discussion about their relevance for control is presented. The paper ends with some conclusions.

\section{Materials and Methods}
The main method is a modulation analysis conducted on each FSMPC variant under consideration: single-vector (SV) and virtual voltage vector (VVV). The analysis is based on the relative usage of each type of basic voltage vectors that the VSI can produce. From this analysis, the values and trends of the figures of merit can be deduced. This helps assessing each FSMPC variant globally (i.e. in all operating regimes of the drive). It is important to note that, most previous comparisons rely on just a handful of operating conditions.

\subsection{Laboratory Setup}
A laboratory system is used for the tests. It includes    a five-phase IM (with  parameters shown in Table \ref{tab_Parameters}), a power converter using two SEMIKRON SKS 22F modules connected to a 300V DC upply, and a MSK28335 system with a TMS320F28335 digital signal processor. The mechanical speed is sensed using a GHM510296R/2500 encoder.  Hall effect sensors (LH25-NP) are used to measure the stator phase currents. Finally, a DC motor is used to generate an opposing torque load ($T_L$) for the tests.

\begin{table}
\centering
\caption{Parameters of the Experimental Five-phase IM.\label{tab_Parameters}}
\begin{tabular}{ccc}
 \textbf{ Parameter} & \textbf{ Value} & \textbf{ Units}  \\\hline
  Stator resistance, $R_s$           &   12.85 &   $\Omega$  \\
 Rotor resistance, $R_r$            &	4.80 &   $\Omega$  \\ 
 Stator leakage inductance, $L_{ls}$&   79.93 &   mH \\
 Rotor leakage inductance, $L_{lr}$ &   79.93 &   mH \\
 Mutual inductance, $L_M$           &  681.7  &   mH \\
 Rotational inertia, $J_m$          &    0.02 &   kg m$^2$ \\
 Number of pairs of poles, $P$      &  3      & - \\ \hline
\end{tabular}
\end{table}

\subsection{Drive control}

\begin{figure}
\centering
\includegraphics[width=83mm]{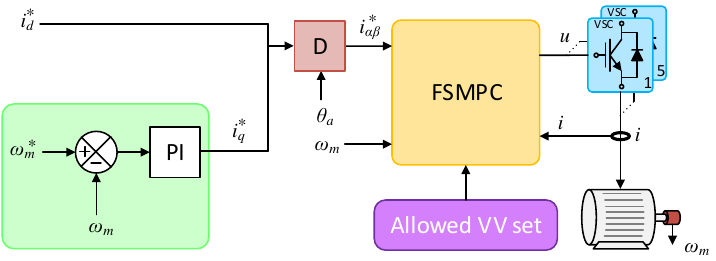}
\caption{Diagram of FSMPC for a five phase IM. \label{fig_diagr}}
\end{figure}

FSMPC for multi-phase drives uses the Clarke projection of stator currents into  the $\alpha-\beta$ and $x_j - y_j$ planes as shown in Figure~\ref{fig_diagr}. For a five-phase VSI, just one $x-y$ plane is produced. The torque producing currents must follow a reference set by the torque/speed control loop. The $x-y$ current references are set to zero to reduce losses. In this way, flux and  torque are independently regulated. The flux set point is provided by $i^{*}_{d}$ whereas reference $i^{*}_{q}$ is used for the electrical torque. These references are projected to  the $\alpha-\beta$ space using the Park transformation, obtaining a reference for stator current in $\alpha-\beta$ plane as $I^*_{s\alpha-\beta} = D \left( i^{*}_{d},  i^{*}_{q} \right)^\intercal$, where matrix $D$ is given by

\begin{equation}\label{eq_park}
  D =
  \begin{pmatrix}
   ~\cos ~ \theta_a  & \sin ~ \theta_a \\
   -\sin ~ \theta_a &  \cos ~ \theta_a\\
  \end{pmatrix}
\end{equation}

The flux angle  $\theta_a$ is obtained as $\theta_a = \int \omega_e dt$. As a result, the set point for stator current tracking $I_s^*(k)$ has an amplitude $I_s^* = \sqrt{  i^{*2}_{d} +  i^{*2}_{q} }$. Finally, the $\alpha-\beta$ references can be expressed as $I^*_{s\alpha}(t)=I_s^* \sin{ \omega_e t}$, $I^*_{s\beta}(t)=I_s^* \cos{ \omega_e t}$, $I^*_{sx}(t)=0$, $I^*_{sy}(t)=0$.

\subsection{VSI States and Voltages}

Phase voltages are set by the VSI state. The state can be represented by a vector $u = \left( K_a, K_b, \cdots , K_e \right)^\top$, where the values $K_{h}$ indicate the state of the corresponding VSI switch for phase $h$. For a five-phase VSI there are 32 configurations. These correspond to values $u_0 = \left( 0, 0, 0, 0, 0 \right)^\intercal$ to $u_{31} = \left( 1, 1, 1, 1, 1\right)^\intercal$. Each state produces a certain set of phase voltages. The stator voltages provided by each VSI state can be found as $V(k) =  V_{DC} T M u(k)$,  where $V_{DC}$ is the  voltage supplying the DC-link and

 \begin{equation}
 T=\frac{1}{5}
 \begin{pmatrix}
 4&-1&-1&-1&-1\\
 -1&4&-1&-1&-1\\
 -1&-1&4&-1&-1\\
 -1&-1&-1&4&-1\\
 -1&-1&-1&-1&4
 \end{pmatrix},
 \end{equation}

 \begin{equation}
 M=\frac{2}{5}
 \begin{pmatrix}
 1&  \gamma^c_1&  \gamma^c_2&  \gamma^c_3&  \gamma^c_4\\
 0&  \gamma^s_1&  \gamma^s_2&  \gamma^s_3&  \gamma^s_4\\
 1&  \gamma^c_2&  \gamma^c_4&  c_\vartheta&  \gamma^c_3\\
 0&  \gamma^s_2&  \gamma^s_4&  \gamma^s_1&  \gamma^s_3\\
 1/2&1/2&1/2&1/2&1/2
 \end{pmatrix}.
 \end{equation}

\noindent where $\gamma^c_h = \cos{h \vartheta}$, $\gamma^s_h = \sin{h \vartheta}$, $\vartheta = {2\pi}/5$ for $h=1, ..., 5$.

\begin{figure}
\centering
\includegraphics[width=83mm]{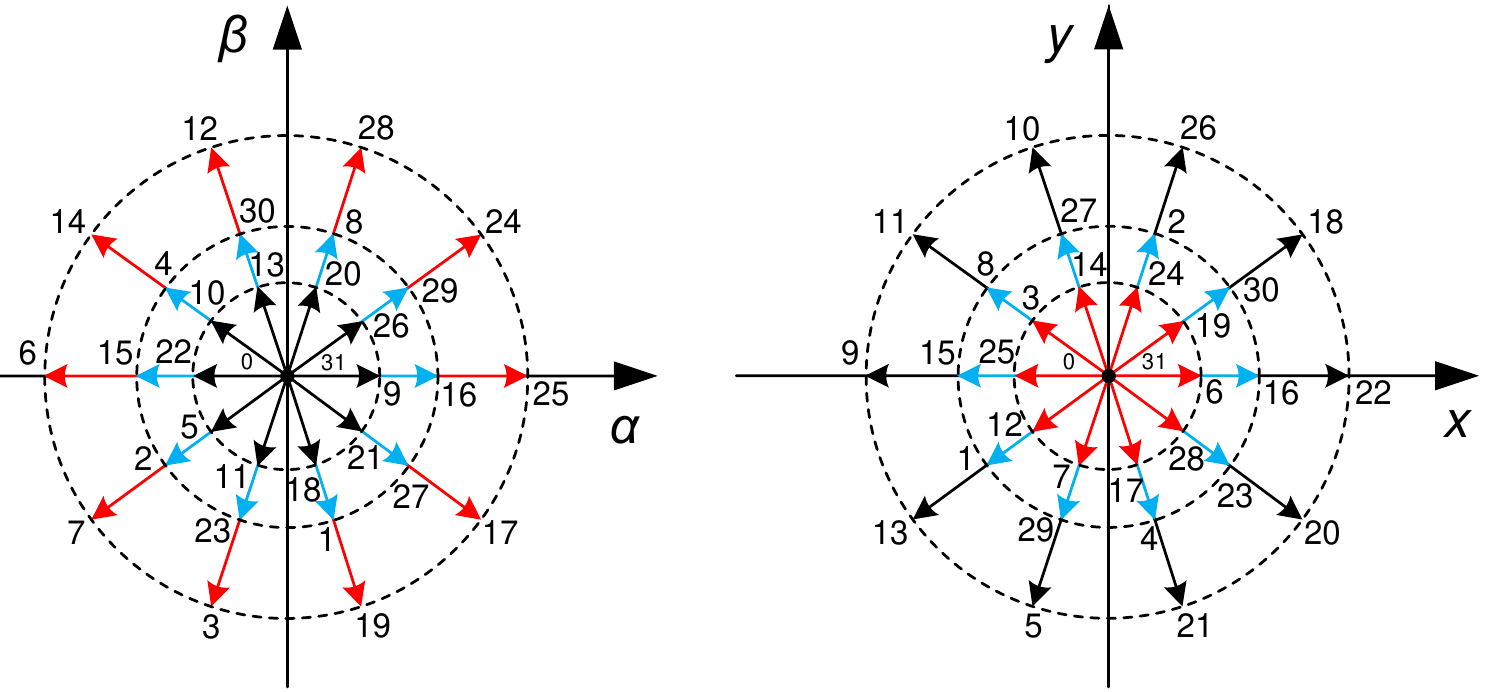}
\caption{Voltage vectors for a five-phase VSI. \label{fig_diagr32VV}}
\end{figure}

The stator voltages can be mapped to $\alpha\beta$ and $xy$ subspaces using the Clarke transformation. The resulting voltages in $\alpha \beta x y$ coordinates are referred to as Voltage Vectors (or VV).  Figure~\ref{fig_diagr32VV} shows the distribution of VV.

It is interesting to describe the distribution attending on the modulus of the VV in the $\alpha\beta$ subspace. This will be useful for the description of the VVV method and for the modulation analysis.

\begin{enumerate}
  \item Large VV. This refers to the VV that have the largest modulus in the $\alpha\beta$ subspace. They lay in the outer corona of the VV distribution in $\alpha\beta$ subspace. These are indicated in Figure~\ref{fig_diagr32VV} with indices: 25, 24, 28, 12, 14, 6, 7, 3, 19, and 17. These VV will be referred to as large VV or just LVV. Please notice that in $xy$ subspace, their modulus is the smallest possible excluding  zero.
  \item Medium VV. This refers to VV with indices 16, 29, 8, 30, 4, 15, 2, 23, 1, and 27. These VV will be referred to as medium VV or just MVV.
  \item Small VV. This refers to VV with indices 9, 26, 20, 13, 10, 22, 5, 11, 18, and 21. These VV will be referred to as small VV or just SVV.
  \item Null voltage. This refers to VV with indices 0, and 31. These VV will be referred to as null VV, zero VV or just ZVV. These configurations produce zero voltage in both subspaces.
\end{enumerate}

\subsection{Predictive Model}

In single-vector PSCC, the control action is the state of the VSI $u$. The PSCC uses a predictive model to link future stator currents to candidate control actions. This model is often a set of discrete-time state-space equations as follows:

\begin{equation}\label{eq_p1p}
\hat{i}(k+1) = \left( C ( \omega) + G \right) i(k) +  B V(k)
\end{equation}

\noindent where  $i$ contains the $\alpha$, $\beta$, $x$ and $y$ stator currents, $\hat{i}$ is the prediction of $i$ and $\omega$ is the angular speed. Matrices $C$ and $B$ are obtained from first principles applying time discretization with sampling time $T_s$. This gives $C = (I + A_c T_s)$, and $B = T_s B_c$, where

\begin{equation}\label{eq_Acont}
A_c =
\begin{pmatrix}
a_{2}& -a_{4}&0&0 \\
a_{4}&  a_{2}&0&0 \\
0&0   & a_{3}&0 \\
0&0&0 & a_{3}
\end{pmatrix},
\end{equation}

\begin{equation}
B_c =
\begin{pmatrix}
c_{2}&0&0&0\\
0&c_{2}&0&0\\
0&0&c_{3}&0\\
0&0&0&c_{3}
\end{pmatrix}.
\end{equation}

Term $G$ accounts for the effect of the rotor currents, which are usually unmeasured variables. In the previous expressions, the coefficients used are:  $a_{2}=-R_{s}c_{2}$, $a_{3}=-R_{s}c_{3}$, $a_{4}=-L_Mc_{4}\omega_{r}$, depending on machine parameters as: $c_{1}=L_{s}L_{r}-L_M^{2}$, $c_{2}=L_{r}/c_{1}$, $c_{3}=1/L_{ls}$, $c_{4}=L_M/c_{1}$. The measured values of the electrical parameters corresponding to the machine used in the experiments are shown in Table \ref{tab_Parameters}.

From equation  (\ref{eq_p1p}) the two-step ahead prediction of the stator currents can be found as

\begin{equation}\label{eq_p2p}
 \hat{i}(k+2) = A ( \omega) \hat{i}(k+1) + B V(k+1)
\end{equation}

\noindent where $A=C+G$ has been introduced for better readability. The predicted control error is thus found as

\begin{equation}\label{eq_error_ctrl_p2p}
 \hat{e}(k+2) = I_s^*(k+2) - \left( A ( \omega) \hat{i}(k+1) + B V(k+1) \right)
\end{equation}

\subsection{Cost function}
The cost function for FSMPC can use different numbers of terms. In this work the following one is used, where the predicted control errors and the VSI commutations are penalized.

\begin{equation}\label{eq_fcoste}
  J(k+2) = \hat{e}_{\alpha-\beta}^2(k+2) + \lambda_{xy} \hat{e}_{x-y}^2(k+2) + \lambda_{sc} \Delta S(k+1),
\end{equation}

where $\Delta S(k+1)$ is the number of switch changes produced at the VSI when configuration $u(k)$ is changed to $u(k+1)$. The value $\Delta S(k+1)$ can be computed as 

\begin{equation}\label{eq_SC}
  \Delta S(k+1) = \sum_{i=1}^{5} | u_i(k+1) - u_i(k) |.
\end{equation}

With this CF structure, the WF are the parameters $\lambda_{xy}$ and $\lambda_{sc}$. The first one allows to put more emphasis on $x-y$ reduction. The second one helps reducing the commutation frequency.

\subsection{Set of Allowed VV}
In single vector FSMPC, the control moves can be drawn from a subset of the possible VSI configurations. This set of allowed control moves is represented by a corresponding set of Allowed VV (AVV). The AVV is often composed of two or more of the following basic VV subsets.

For the purposes of this paper two choices of AVV set are considered: Large plus Zero and Full. These are described below.

\begin{enumerate}
  \item $\zeta_L$. This group of VVs is composed of the LVV set and the ZVV set. In other words, it contains the 10 LVV plus the zero VV, so $\zeta_L = LVV  \cup  ZVV$.

  \item $\zeta_{W}$. Whole set. This group of VVs is composed of the LVV set, the MVV set, the SVV set and the ZVV set.
\end{enumerate}

\subsection{Virtual VV}
The so called Virtual Voltage Vectors method also makes use of VV from different subsets: LVV, MVV, etc. However this approach is not single vector, meaning that more than one VV is issued per sampling period.

Although there are some variants of the VVV technique, the basic idea is  to issue two VVs during a single control period. The VVs are selected so that their average $x-y$ excitation is zero (in the ideal case). In this way the terms penalizing $x-y$ currents in the cost function can be dropped. This allows the commutation term to be dropped from the cost function. On top of that, the number of VVV is just 11 (for the five-phase VSI), instead of the 32 VVs.

For the five-phase VSI, the VVV are formed by issuing a Large VV during a subperiod $T^L$ followed by a Medium VV during $T^M$. The values $T^L$ and $T^M$ are chosen so that they add up to the sampling period, this means $T^L+T^M=T_s$. Also, the average voltage (over a sampling period) in $x-y$ subspace is made zero: $\overline{V}_{xy}=0$. This average voltage is found as

\begin{equation}\label{eq_averageVxy}
  \overline{V}_{xy} = \frac{ V^L_{xy} \cdot T^L + V^M_{xy} \cdot T^M }{T_s},
\end{equation}

\noindent where $V^L_{xy}$ is the modulus of any Large VV in the $x-y$ subspace and $V^M_{xy}$ is the modulus of any Medium VV in the $x-y$ subspace. The cancellation is possible by pairing a LVV (such as 20) with the one MVV lying in the opposite part of the $x-y$ subspace (such as 8). This produces $T^L = 0.618 T_s$, $T^M = 0.382 T_s$.

The VVV formed in this way are referred to as L+M VVV.

\subsection{Figures of Merit}
A set of figures of merit is used to assess the different FSMPC variants. These are described in the following.

\begin{enumerate}
    \item Tracking error in $\alpha-\beta$ subspace ($E_{\alpha-\beta}$). This indicator is important to define the quality of stator currents. It is also linked to torque ripple.
    \item Regulating error in $x-y$ subspace ($E_{xy}$). This error is important for energy efficiency as   $x-y$ currents produce losses.
    \item Average Switching Frequency (ASF). The switching frequency (for most FS-MPC methods) is not constant. The average value has importance for VSI commutation losses.
    \item Voltage Total Harmonic Distortion (THD). This indicator is of importance for grid quality as harmonics often  have negative effects.
\end{enumerate}

The above indicators are defined as follows.

\begin{eqnarray} \label{eq_def_eab}
  E_{\alpha-\beta} &=& \sqrt{ \frac{1}{(k_2-k_1+1)} \sum_{k=k_1}^{k_2} e_{\alpha\beta}^2 (k) } \\ \label{eq_def_exy}
  E_{x-y}          &=& \sqrt{ \frac{1}{(k_2-k_1+1)} \sum_{k=k_1}^{k_2} e_{xy}^2 (k) } \\ \label{eq_def_frec_sw}
  ASF &=& \frac{1/(5 \cdot 2)}{t(k_2) -t(k_1)} \sum_{k=k_1}^{k_2} \Delta S(k) \\
   \label{eq_def_thd}
  THD_V &=& \frac{100}{V_1} \sqrt{ \sum_{i=2}^{\infty} V_i^2 }
\end{eqnarray}

\noindent where $\Delta S(k) = \sum_{i=1}^{5} | u_i(k+1) - u_i(k) |$ is the number of switch changes produced at the VSI when configuration $u(k)$ is changed to $u(k+1)$, and $V_i$ is the amplitude of the $i$-th harmonic component of stator voltages. These quantities are defined over a temporal horizon defined by the discrete-time indices $k_1$, $k_2$, for which $t( k) = k\cdot T_s$ with $T_s$ the sampling period of FSMPC. Please notice that for the VVV method, the inter-sample VSI commutations must be counted as well. These correspond to switching from a L VVV to an M VVV and amount to 2 changes except for the special case of the null VVV where the inter-sample changes are zero.

\subsection{Modulation Analysis}
\begin{figure}
  \centering
    \includegraphics[width=8cm]{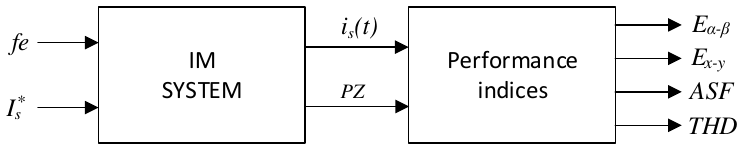} \\
    \caption{Performance indices.}\label{fig_performance_indices}
\end{figure}

The modulation analysis is illustrated by the diagram of Figure \ref{fig_performance_indices}. The study uses electrical frequency $f_e$ and amplitude of stator current reference $I_s^*$ as fundamental variables to explore the operating regimes of the multi-phase drive. In this way the actual speed-torque curve of any specific passive mechanical load is not needed. Instead, a broad set of conditions are tested, providing a global picture of the drive behavior. 

As indicated in Figure \ref{fig_performance_indices}, for each $f_e$ and $I_s^*$ the closed loop operation of the drive is considered. Because of the FSMPC scheme, the sequence of VV is intended to produce a sinusoidal current in $\alpha-\beta$ subspace with amplitude $I_s^*$ and  frequency $f_e$. Then, the occurrence of the various VV is investigated. In particular the usage of the ZVV is recorded as it plays a big role in drive behavior, in particular in ASF. Finally, the resulting stator current can be characterized using the figures of merit already presented.

\begin{figure}
  \centering
  \begin{tabular}{c}
\includegraphics[width=8cm]{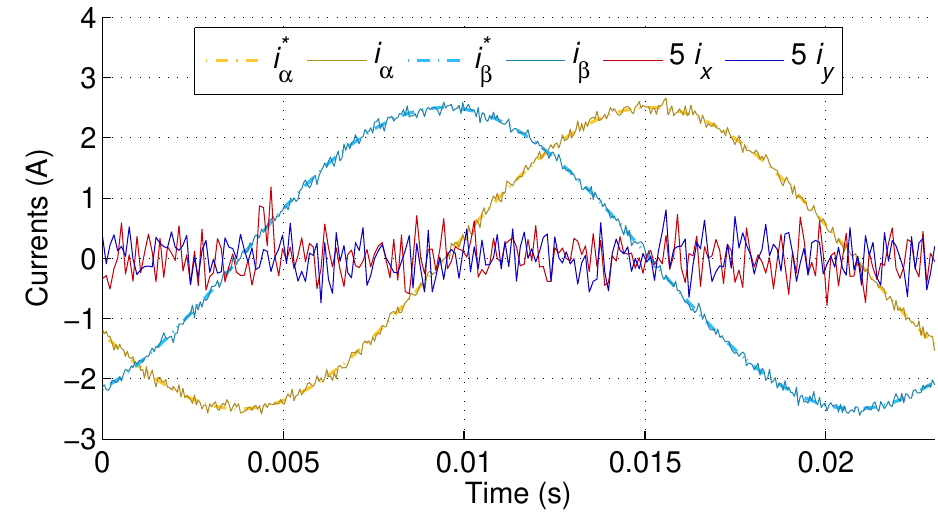} \\
   a) M+L VVV  \\
  \includegraphics[width=8cm]{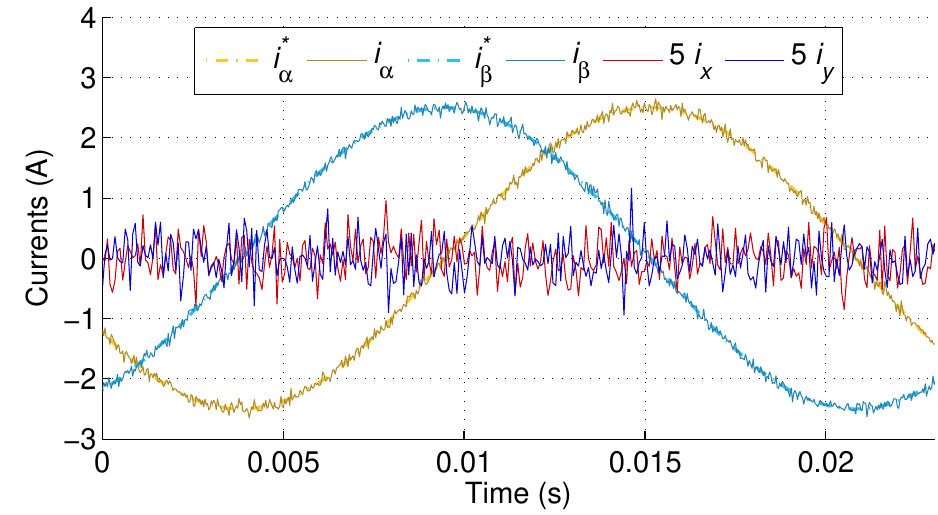} \\
   b) SV 32VV WF 
  \end{tabular}
  \caption{Evolution of stator currents for some FSMPC variants for rated speed and load.}\label{fig_trayect_casos}
\end{figure}

As an example consider Fig. \ref{fig_trayect_casos} where the trajectories of stator currents are shown for some FSMPC variants. The figures of merit are derived from these trajectories to obtain some conclusions. The FSMPC variants in Fig. \ref{fig_trayect_casos} are: a) [M+L VVV] multi-vector scheme using the L+M set of virtual VV, and b) [SV 32VV WF] single-vector using $\zeta_W$ as AVV set and WF.

\section{Results}
The analysis described in the previous section is now carried out for different FSMPC variants using a multi-phase VSI.  The first variant is SV-FSMPC with a reduced AVV set.

The $\zeta_L$ set of VVs is  the simplest choice for a reduced AVV set. It has been promoted in some papers as a means to deal with the complexity of CF optimization. Here  $\zeta_L$ is used mainly as a stepping stone for more complex analysis. Yet, even in this simple case, new observations arise that are worth contemplating. Also, this case will provide a base-line for the comparison with multi-vector VVV approach.

\subsection{SV-FSMPC Using Large Vectors as AVV Set}
The analysis of SV-FSMPC using Large Vectors as AVV set produces the results of Figure~\ref{fig_res_10LPZ}. For this case, values $\lambda_{xy}=0$ and $\lambda_{sc}=0$ are used. The effect of WF is to be discussed later on.

The first row (a) of Figure~\ref{fig_res_10LPZ} shows the usage of the ZVV for different operating regimes. This set of curves expose the implicit modulation taking place in SV-FSMPC. It will be shown that the $PZ$ curves allows explaining many characteristics of SV-FSMPC behavior. In particular, the figures of merit (of rows b) to e)) are mostly a consequence of the modulation. The analysis relies on the following observations.

\subsubsection{Observations}
\begin{figure}
\centering
\begin{tabular}{m{0.25cm} m{8cm}}
 {\scriptsize a)} & \includegraphics[width=80mm]{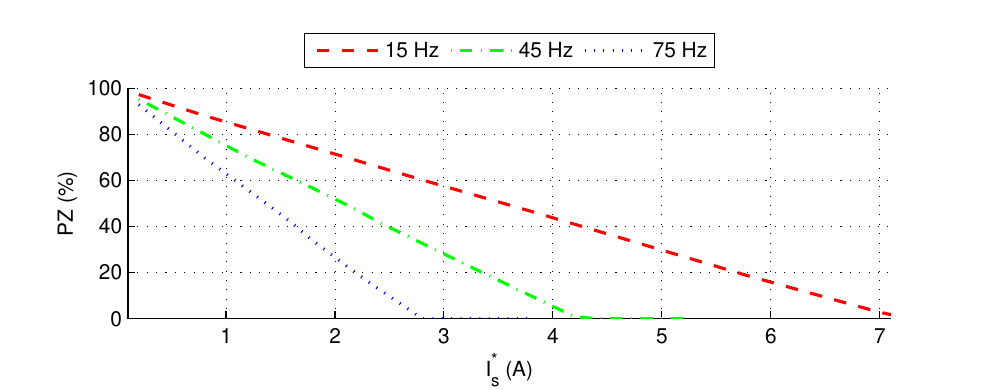} \\
 {\scriptsize b)} & \includegraphics[width=80mm]{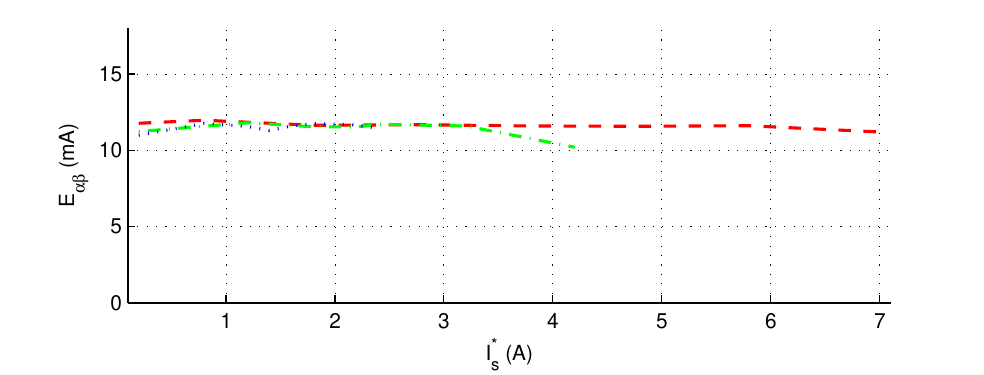} \\
 {\scriptsize c)} & \includegraphics[width=80mm]{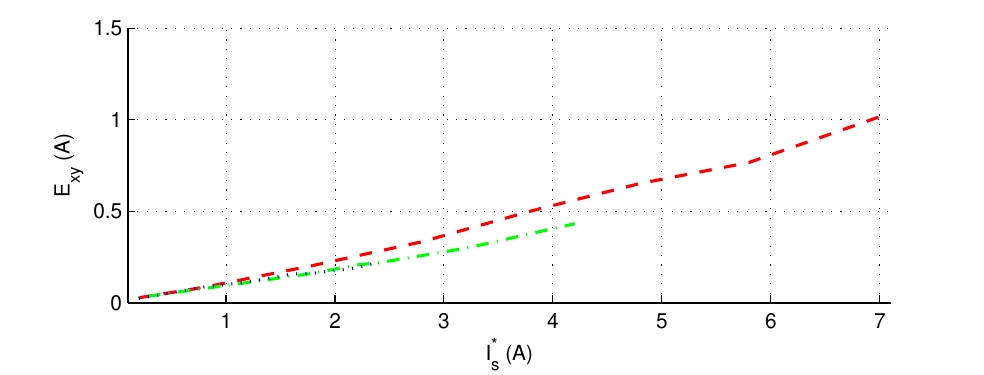} \\
 {\scriptsize d)} & \includegraphics[width=80mm]{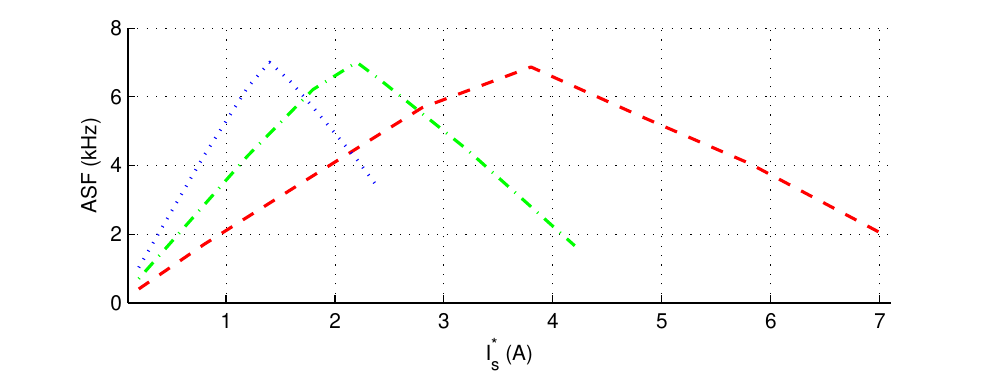} \\
 {\scriptsize e)} & \includegraphics[width=80mm]{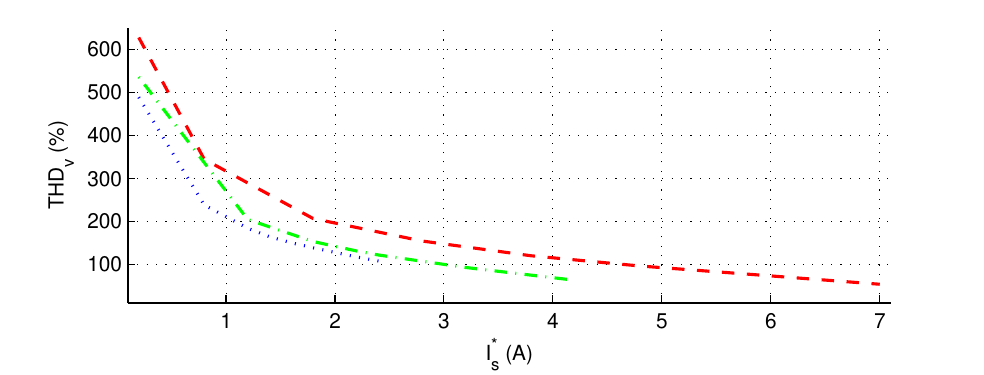}
\end{tabular}
  \caption{Figures of merit of SV-FSMPC with $AVV=\zeta_{L}$. Each curve correspond to a $f_e$ value as indicated by the legend at the top.}\label{fig_res_10LPZ}
\end{figure}

Considering Figure~\ref{fig_res_10LPZ} and proceeding from top to bottom.

\begin{enumerate}
  \item[O1] From row a), $PZ$ diminishes steadily with $I_s^*$. This is a result of the need to reduce the effective (average) voltage caused by the LVV. This is achieved by SV-FSMPC by using the ZVV. The higher $I_s^*$ the higher the needed voltage and the lower the ZVV usage.
  \item[O2] In row b), the control error  ($E_{\alpha\beta}$) shows little variation with both $I_s^*$ and $f_e$. This is the result of the CF penalizing predicted deviations of $I_{\alpha\beta}$ from its reference.
  
  \item[O3] From row c), $E_{xy}$ grows with $I_s^*$. This is a consequence of the reduction in $PZ$ (see O1). Recall that the LVV excite the $x-y$ subspace whereas the ZVV do not. As a result, the lower $PZ$ the higher  $E_{xy}$. Also recall that, in this test, $\lambda_{xy}=0$. This means that the controller makes no effort in reducing $x-y$ content.

  \item[O4] In row d), $ASF$ is observed to exhibit a non-linear relationship with $I_s^*$. This has been overlooked in many papers although it is a simple consequence of modulation. For low $I_s^*$, consecutive samplings often use a ZVV, needing no commutations, hence one gets a low ASF. For high $I_s^*$ values, consecutive samplings often the same LVV (or adjacent ones) again needing a low number of commutations. Finally, for intermediate values of $PZ$, consecutive samplings often switch from ZVV to a LVV, needing 2 or 3 leg switches. 
  
  \item[O5] From row e), $THD_v$ exhibit quite large values for low $I_s^*$ that are seldom reported. The larger values $THD_v$ appear for larger $PZ$. The reason is that the ZVV reduce the fundamental component of the waveform ($V_1$). This makes the harmonic to $V_1$ ratio higher.
  
  \item[O6] For larger $f_e$ the range of attainable $I_s^*$ is smaller due to the limitation placed by the DC-link voltage. This can be seen in Figure~\ref{fig_res_10LPZ} a) where larger $f_e$ values need less ZVV and use the LVV sooner. This has the effect of shifting the curves to the left (lower  $I_s^*$) as $f_e$ increases.

\end{enumerate}

\subsubsection{Relevance for control}
The above observations are useful as pieces of evidence to assess this SV-FSMPC variant and to compare with others.

\begin{enumerate}
  \item O1 shows that SV-FSMPC has a built-in, flexible modulation capability. The flexibility is seen as different $I_s^*$ and different $f_e$ values are handled to attain the objective (in this case $\alpha\beta$ tracking). This observation reduces the theoretical support for schemes that introduce explicit, a priori, hard-wired modulations such as VVV.
  
  \item O2 shows that SV-FSMPC is able to produce accurate tracking of the torque-producing stator currents for different conditions.
  
  \item O3 supports the use of WF or other methods for $E_{xy}$ reduction. Also it prompts for methods  that can handle different operating regimes as the unconstrained $E_{xy}$ content varies with $I_s^*$ and  $f_e$ .
  
  \item O4 provides theoretical support for the often reported idea that $ASF$ is smaller in SV-FSMPC than in PWM (when using a carrier frequency equal to $1/T_s$ (Hz)). The observation can even be used to estimate the maximum ASF for a given $T_s$. Since 2-3 commutations are expected in the worst case, the  ratio $2.5/(5 \cdot 2 \cdot T_s)$ is an estimate of the ASF in the zone where $PZ \approx 50$ \%. In the tests this amounts to an estimate of 7.1 kHz that is in good accordance with the observed value.
  
  O4 also shows that switching penalization methods must consider the operating regime, which is often overlooked.
  
  \item O5 show that $\zeta_L$ is a bad choice in cases where $THD_V$ is of importance, specially in the low load region. This fact is overlooked quite often. This is relevant since other schemes are similar to $\zeta_L$ in this regard.
  
  \item O6 support the need to consider multiple operating regimes for the assessment of FSMPC for drives as the results for some figures of merit are quite different.
\end{enumerate}

\subsection{MV-FSMPC Using M+L VVV}
\begin{figure}
\centering
\begin{tabular}{m{0.25cm} m{8cm}}
 {\scriptsize a)} & \includegraphics[width=80mm]{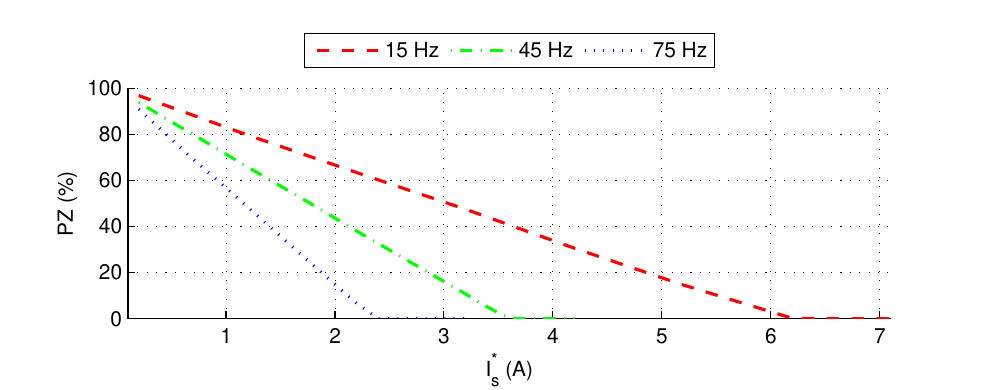} \\
 {\scriptsize b)} & \includegraphics[width=80mm]{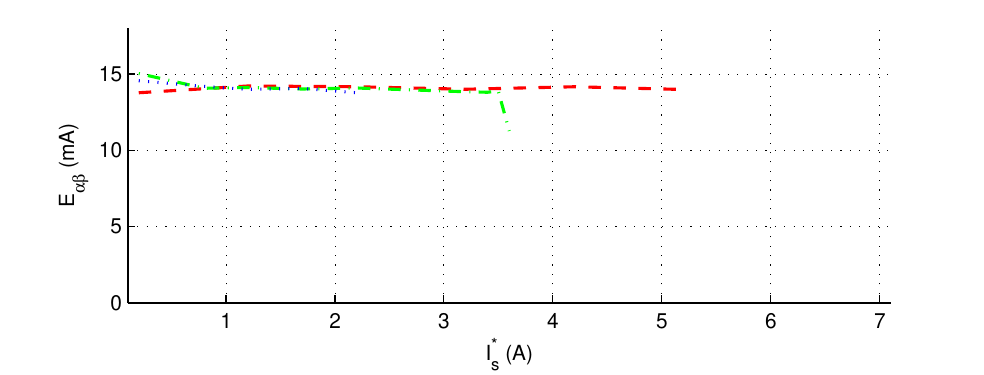} \\
 {\scriptsize c)} & \includegraphics[width=80mm]{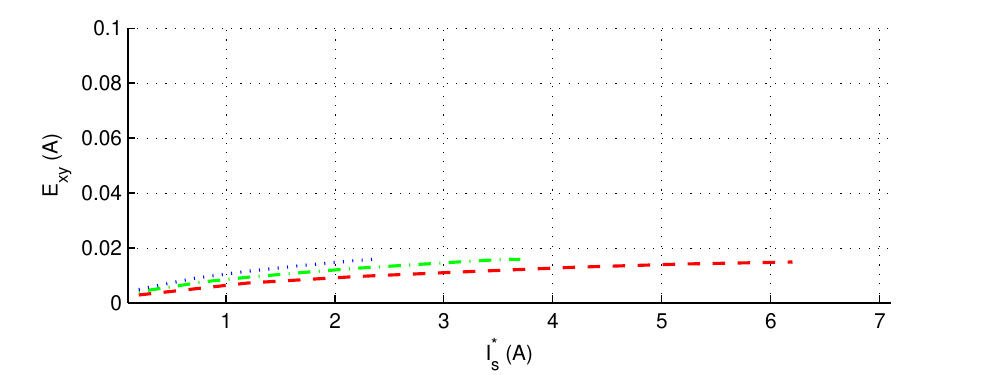} \\
 {\scriptsize d)} & \includegraphics[width=80mm]{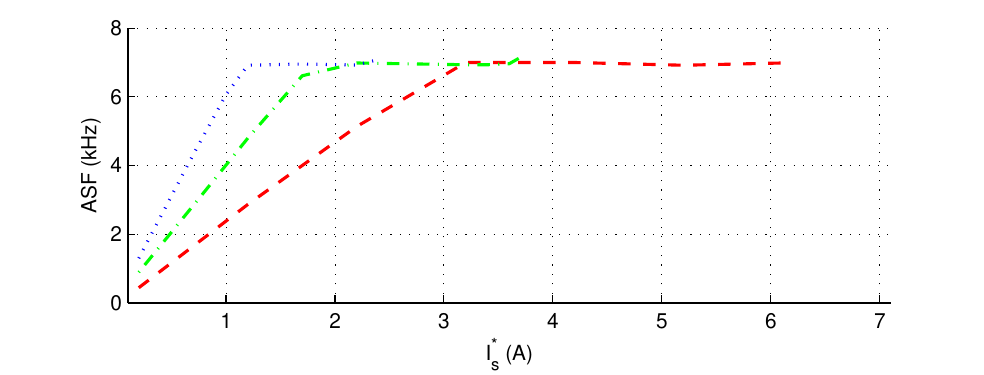} \\
 {\scriptsize e)} & \includegraphics[width=80mm]{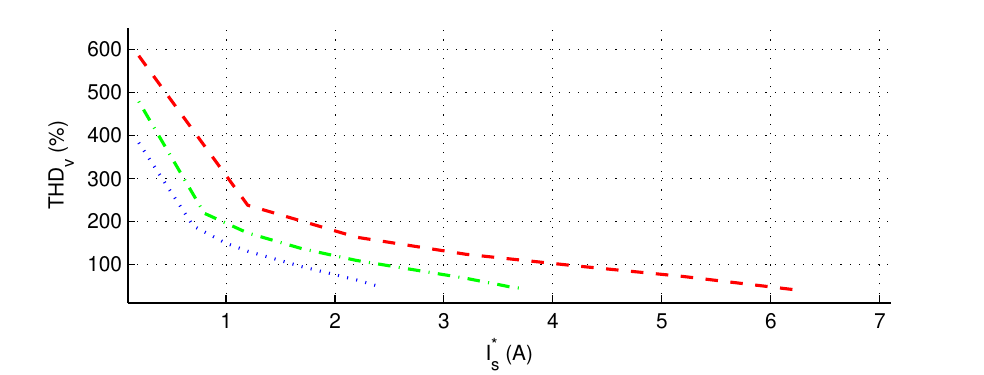}
\end{tabular}
  \caption{Results for multi-vector M+L VVV. Each curve correspond to a $f_e$ value as indicated by the legend at the top. Note the change in scale for $E_{xy}$ with respect to the SV-FSMPC $\zeta_L$ case.}\label{fig_res_VVV_ML}
\end{figure}

A multi-vector FSMPC variant is now analyzed. The technique corresponds to VVV using medium and large VV. The results are presented in Figure~\ref{fig_res_VVV_ML}. Since this technique uses a single corona (formed by the VVV), it is interesting to draw a comparison with the case of SV-FSMPC using  a single corona of basic VVs, for instance $\zeta_L$.

\subsubsection{Observations}
The following observations can be made on Figure~\ref{fig_res_VVV_ML}.

\begin{enumerate}
  \item[O1] $PZ$ diminishes steadily with $I_s^*$. The usage is lower than in the case of SV-FSMPC with $\zeta_L$ as AVV set. This is a result of the lower modulus of the VVV compared with the modulus of LVV. This means that the $PZ$ curves are shifted to the left compared with the $\zeta_L$ case. This will produce a similar shift for the rest of figures of merit.
  \item[O2] $E_{\alpha\beta}$ is a bit higher than in the SV-FSMPC case. Also the  maximum attainable $I_s^*$ is smaller.
  \item[O3] The $E_{xy}$ values are quite low. Recall that this result is for an ideal case, so that the open loop $xy$ regulation works almost perfectly.
  \item[O4] $ASF$ values are higher than in the $\zeta_L$ case. This is caused by the fact that commuting from an M VV to an L VV always needs 2 switch changes. 
  
  \item[O5] $THD_v$ is smaller than in the $\zeta_L$ case due to the reduced use of the ZVV.
\end{enumerate}

The above observations support the following assessment of M+L VVV compared with SV-FSMPC using $\zeta_L$ as the AVV set: just $E_{xy}$ and $THD_v$ improve, the other figures of merit get worse. In particular the DC-bus usage is reduced.

\subsection{SV-FSMPC with LVV and Fixed Weighting Factors}
\begin{figure}
\centering
\begin{tabular}{ll}
 {\footnotesize a)} & \includegraphics[width=80mm]{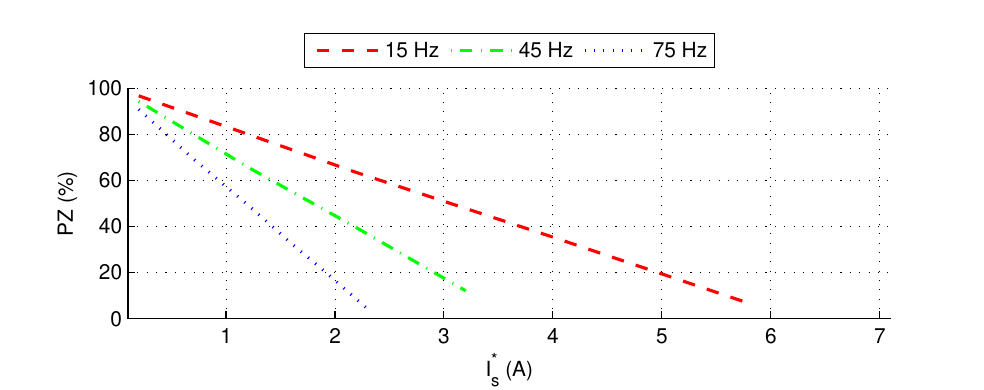} \\
 {\footnotesize b)} & \includegraphics[width=80mm]{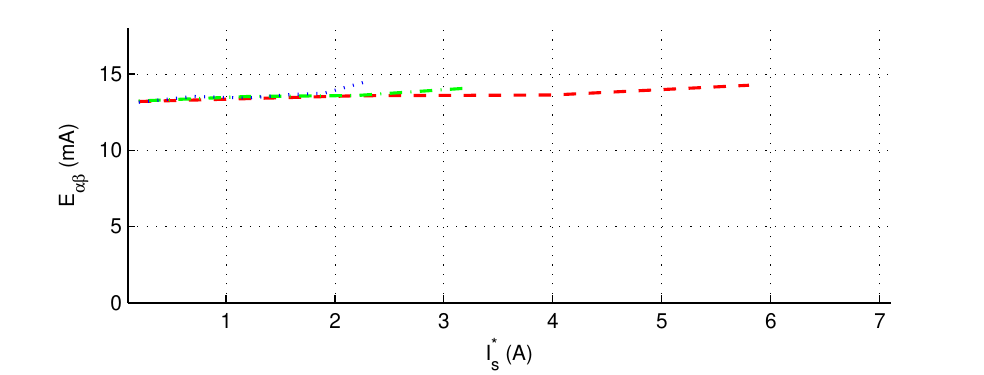} \\
 {\footnotesize c)} & \includegraphics[width=80mm]{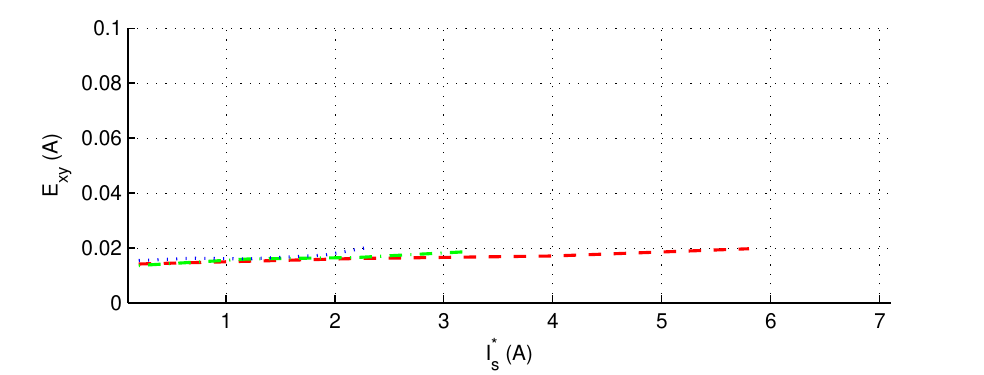} \\
 {\footnotesize d)} & \includegraphics[width=80mm]{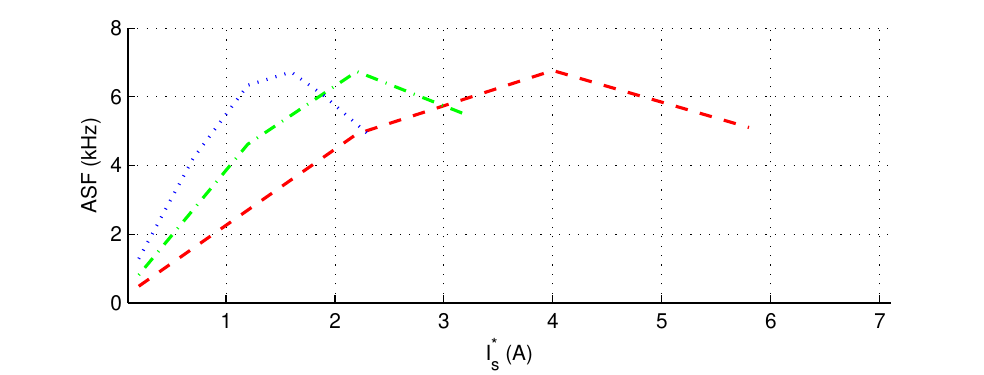} \\
 {\footnotesize e)} & \includegraphics[width=80mm]{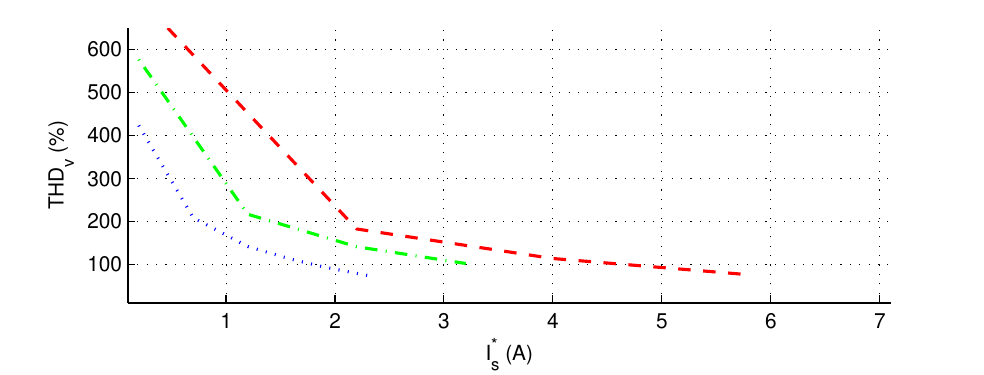}
\end{tabular}
  \caption{Figures of merit of SV-FSMPC for $\zeta_L$ using $\lambda_{xy}=0.5$ and $\lambda_{sc}=0$. Each curve correspond to a $f_e$ value as indicated by the legend at the top.}\label{fig_res_10LPZWF}
\end{figure}

The single-vector case is re-evaluated, this time making use of WF. The rationale for this is double. First, to establish whether or not SV-FSMPC can provide the low values of $E_{xy}$ observed in M+L VVV. Second, to illustrate the flexibility provided by the WF. Figure~\ref{fig_res_10LPZWF} presents the results for $\lambda_{xy}=0.5$ and $\lambda_{sc}=0$.

It can be seen that all figures of merit are very similar to the VVV case. The values of the WF have been adjusted so that the ASF for both approaches is almost the same. In fact it is a bit higher for the M+L VVV technique, giving it a bit of advantage. The DC-link usage is better for the single-vector approach, as expected.

\subsection{SV-FSMPC with Full VV Set and Fixed Weighting Factors}
The reduced set case has been useful to provide a straight comparison of a SV-FSMPC method with a multi-vector one (M+L VVV). However, unlike the VVV technique, the SV-FSMPC does not need to be restricted to the use of a single corona of VV. This is more so  after the recent introduction of region-based methods enabling a fast computation of the control signal in SV-FSMPC. 

\begin{figure}
\centering
\begin{tabular}{ll}
 {\footnotesize a)} & \includegraphics[width=80mm]{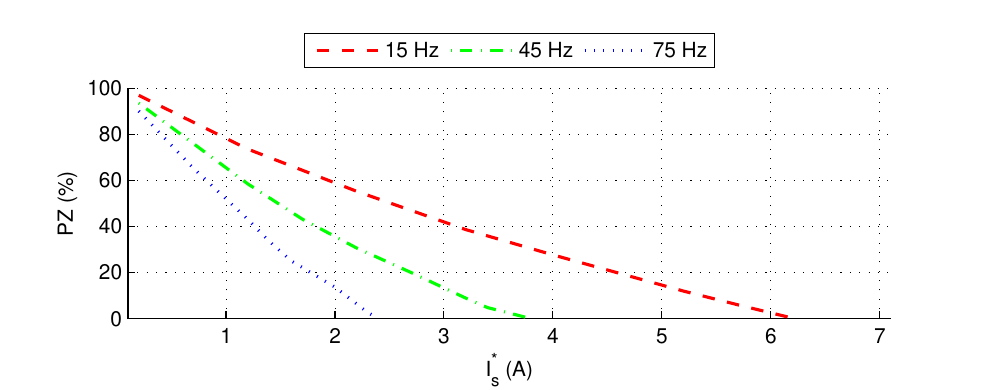} \\
 {\footnotesize b)} & \includegraphics[width=80mm]{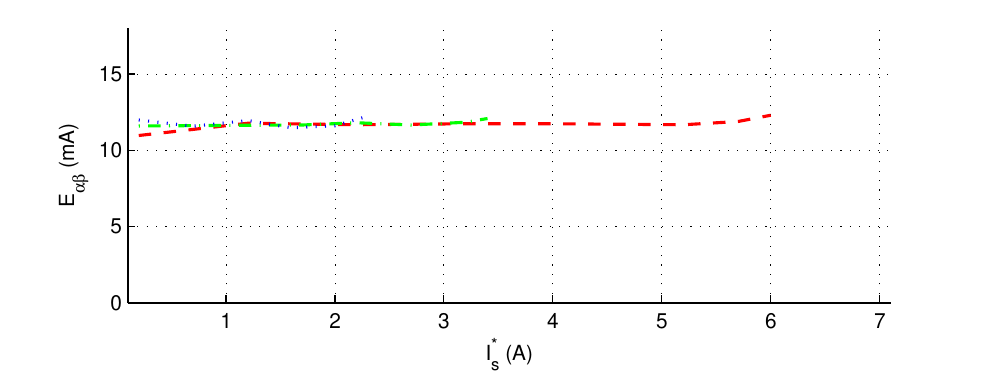} \\
 {\footnotesize c)} & \includegraphics[width=80mm]{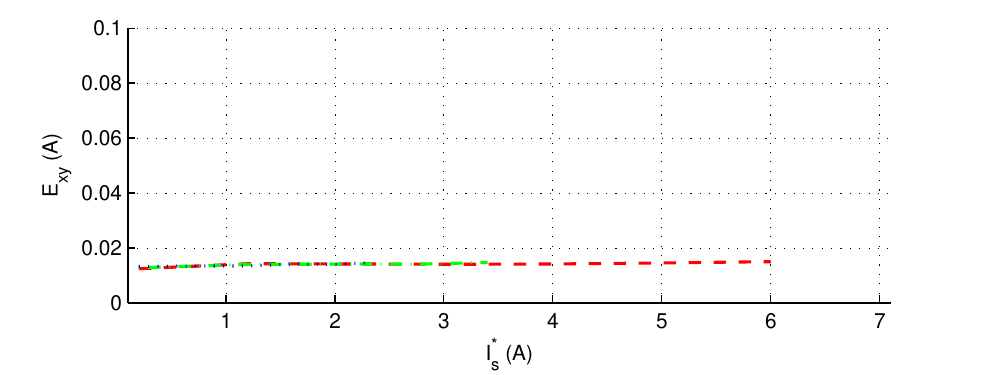} \\
 {\footnotesize d)} & \includegraphics[width=80mm]{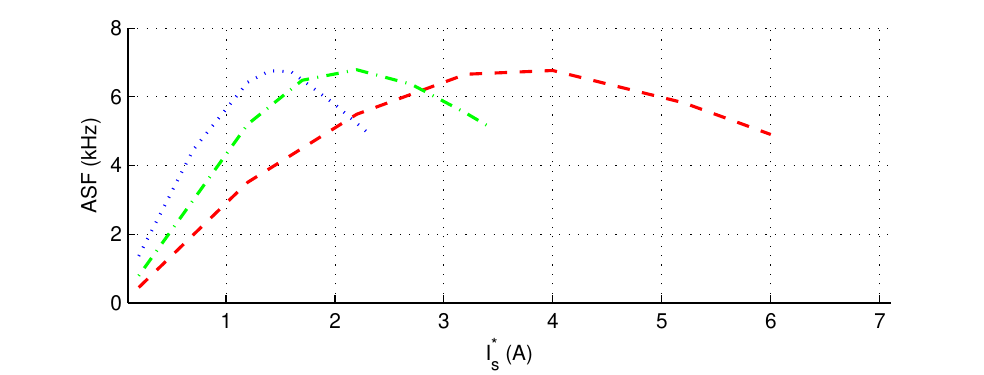} \\
 {\footnotesize e)} & \includegraphics[width=80mm]{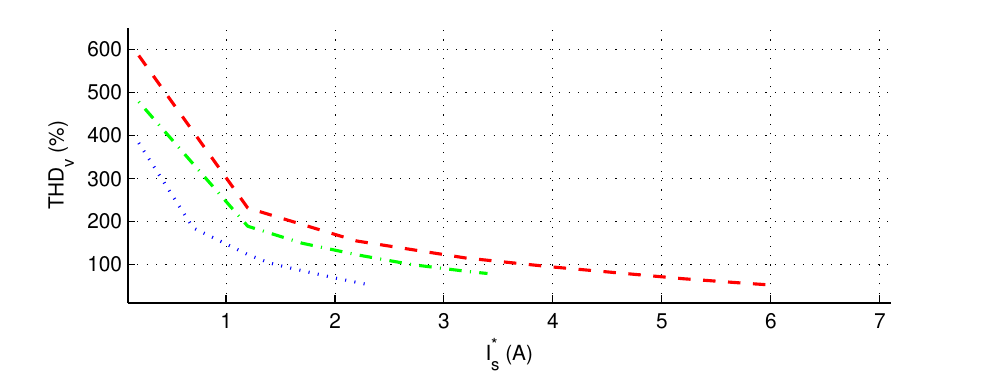}
\end{tabular}
  \caption{Figures of merit of SV-FSMPC for $\zeta_W$ using $\lambda_{xy}=0.72$ and $\lambda_{sc}=0$. Each curve correspond to a $f_e$ value as indicated by the legend at the top.}\label{fig_res_32VVWF}
\end{figure}

In the following, the full set of VVs is considered. In other words, the AVV set is $\zeta_W$. The region-based method of \cite{arahal2024multi} is used to compute the control action. The results are shown in  Figure~\ref{fig_res_32VVWF}, where the WF used are $\lambda_{xy}=0.72$ and $\lambda_{sc}=0$.

Compared with the L+M VVV approach,  all figures of merit are very similar. The values of the WF have been adjusted so that the ASF for both approaches is almost the same. In fact it is a bit higher for the M+L VVV technique, giving it a bit of advantage. The DC-link usage is better for the single-vector approach. Better $E_{\alpha-\beta}$ values can be obtained with SV-FSMPC without increasing much the $x-y$ content and for slightly less ASF than L+M VVV. Finally, the THD values are the lowest of all the techniques considered so far.

\section{Discussion}

The M+L VVV scheme uses a virtual set of vectors that resemble the $\zeta_L$ case. It does not  use WF because both ASF and $E_{xy}$ are  acceptable for many applications. It benefits from a reduction in CF complexity. However, it can be argued that the $\zeta_L$ case with WF could compete with M+L VVV.

From the analysis it seems clear that M+L VVV is appropriate for $E_{xy}$ reduction at the cost of higher ASF. In a practical situation the DC-link usage reduction can be an important drawback. Another issue for the VVV technique is the fact that $xy$ regulation is done in open loop.

At the moment of its introduction, the VVV technique had the advantage of simplifying the CF. This, in turn, was used to diminish the computational load, so allowing a reduction in the sampling period. Today, however, fast computation of the control action is possible for single vector FSMPC using WF \cite{arahal2024multi}. 

The SV-FSMPC also provides flexibility to adjust the figures of merit. For instance, in some applications one might want to trade some $E_{xy}$ for a better $\alpha-\beta$ tracking. The results obtained so far support the idea of balancing figures of merit with a broader picture. This new view is provided by the use of a broad range of loads and speeds corresponding to different values of $f_e$ and $I_s^*$. To support this assertion just compare the results of Figure~\ref{fig_res_10LPZ} (obtained for $\lambda_{xy}=0$, $\lambda_{sc}=0$) with the results of Figure~\ref{fig_res_10LPZWF} where $\lambda_{xy}=0.5$. This flexibility is lacking in the multi-vector approaches.


\section{Acknowledgments}

\bibliographystyle{IEEEtran}
\bibliography{00main}

\begin{thebibliography}{10}
\providecommand{\url}[1]{#1}
\csname url@samestyle\endcsname
\providecommand{\newblock}{\relax}
\providecommand{\bibinfo}[2]{#2}
\providecommand{\BIBentrySTDinterwordspacing}{\spaceskip=0pt\relax}
\providecommand{\BIBentryALTinterwordstretchfactor}{4}
\providecommand{\BIBentryALTinterwordspacing}{\spaceskip=\fontdimen2\font plus
\BIBentryALTinterwordstretchfactor\fontdimen3\font minus \fontdimen4\font\relax}
\providecommand{\BIBforeignlanguage}[2]{{%
\expandafter\ifx\csname l@#1\endcsname\relax
\typeout{** WARNING: IEEEtran.bst: No hyphenation pattern has been}%
\typeout{** loaded for the language `#1'. Using the pattern for}%
\typeout{** the default language instead.}%
\else
\language=\csname l@#1\endcsname
\fi
#2}}
\providecommand{\BIBdecl}{\relax}
\BIBdecl

\bibitem{kennel2000predictive}
R.~Kennel and A.~Linder, ``Predictive control of inverter supplied electrical drives,'' in \emph{2000 IEEE 31st Annual Power Electronics Specialists Conference. Conference Proceedings (Cat. No. 00CH37018)}, vol.~2.\hskip 1em plus 0.5em minus 0.4em\relax IEEE, 2000, pp. 761--766.

\bibitem{lim2022continuous}
C.~S. Lim, S.~S. Lee, and E.~Levi, ``Continuous-control-set model predictive current control of asymmetrical six-phase drives considering system nonidealities,'' \emph{IEEE Transactions on Industrial Electronics}, vol.~70, no.~8, pp. 7615--7626, 2022.

\bibitem{an2024robust}
X.~An, Z.~Liu, Q.~Chen, and G.~Liu, ``Robust virtual-vector model predictive control of permanent-magnet motor considering {D-Q} axis inductance parameter uncertainty,'' \emph{IET Electric Power Applications}, vol.~18, no.~1, pp. 76--89, 2024.

\bibitem{liu2024sensorless}
T.~Liu, Q.~Zhao, K.~Zhao, L.~Li, and G.~Zhu, ``Sensorless model predictive control of permanent magnet synchronous motor based on hybrid parallel observer under parameter uncertainty,'' \emph{IET Power Electronics}, vol.~17, no.~3, pp. 438--449, 2024.

\bibitem{alharbi2024review}
A.~Alharbi, S.~Odhano, A.~Smith, X.~Deng, and B.~Mecrow, ``A review of modeling and control of multi-phase induction motors under machine faults,'' in \emph{2024 International Conference on Electrical Machines (ICEM)}.\hskip 1em plus 0.5em minus 0.4em\relax IEEE, 2024, pp. 1--9.

\bibitem{taherzadeh2024six}
M.~Taherzadeh, H.~H{\'e}nao, and G.-A. Capolino, ``Six-phase induction machines: State of the art on design, modeling, control and diagnosis,'' in \emph{2024 International Conference on Electrical Machines (ICEM)}.\hskip 1em plus 0.5em minus 0.4em\relax IEEE, 2024, pp. 1--7.

\bibitem{borreggine2019review}
S.~Borreggine, V.~G. Monopoli, G.~Rizzello, D.~Naso, F.~Cupertino, and R.~Consoletti, ``A review on model predictive control and its applications in power electronics,'' in \emph{2019 AEIT International Conference of Electrical and Electronic Technologies for Automotive (AEIT AUTOMOTIVE)}.\hskip 1em plus 0.5em minus 0.4em\relax IEEE, 2019, pp. 1--6.

\bibitem{xue2023recent}
Z.~Xue, S.~Niu, A.~M.~H. Chau, Y.~Luo, H.~Lin, and X.~Li, ``Recent advances in multi-phase electric drives model predictive control in renewable energy application: A state-of-the-art review,'' \emph{World Electric Vehicle Journal}, vol.~14, no.~2, p.~44, 2023.

\bibitem{mamdouh2022simple}
M.~Mamdouh and M.~A. Abido, ``Simple predictive current control of asymmetrical six-phase induction motor with improved performance,'' \emph{IEEE Transactions on Industrial Electronics}, vol.~70, no.~8, pp. 7580--7590, 2022.

\bibitem{serra2021computationally}
J.~Serra, I.~Jlassi, and A.~J.~M. Cardoso, ``A computationally efficient model predictive control of six-phase induction machines based on deadbeat control,'' \emph{Machines}, vol.~9, no.~12, p. 306, 2021.

\bibitem{arahal2024multi}
M.~Arahal, M.~Satu{\'e}, and F.~Barrero, ``Multi-phase weighted stator current tracking using a hyper-plane partition of the control set,'' \emph{Control Engineering Practice}, vol. 153, p. 106114, 2024.

\bibitem{he2024model}
J.~He, Z.~Li, G.~Wu, and R.~Tang, ``Model predictive stator flux control of permanent magnet synchronous motor based on vector duty ratio modulation,'' \emph{IEEE Access}, 2024.

\bibitem{li2024novel}
M.~Li, J.~Zhu, Q.~Liu, H.~Liao, and K.~Zang, ``A novel model predictive current control for fault tolerant permanent magnet vernier rim-driven motor based on improved sector selection,'' \emph{Journal of Electrical Engineering \& Technology}, pp. 1--10, 2024.

\bibitem{liu2024three}
Y.~Liu, S.~Huang, W.~Liao, G.~Liang, H.~Cui, C.~Feng, X.~Wu, J.~Wei, and S.~Huang, ``Three-vector-based model predictive torque control for dual three-phase {PMSM} with torque and flux ripples reduction,'' \emph{IEEE Transactions on Power Electronics}, 2024.

\bibitem{juan12guiding}
J.~Juan, G.~Ignacio, J.~Mario, G.~Angel, and C.~Juan, ``Guiding the selection of multi-vector model predictive control techniques for multiphase drives,'' \emph{Machines}, vol.~12, no.~2, p. 115, 2024.

\bibitem{yan2024synthetic}
L.~Yan, X.~Zhang, J.~Yang, G.~Yang, and R.~Deng, ``Synthetic vectors-based predictive control of dual three-phase {PMSM}s for current harmonics mitigation considering average deception effect,'' \emph{IEEE Journal of Emerging and Selected Topics in Power Electronics}, 2024.

\bibitem{arahal2016harmonic}
M.~R. Arahal, F.~Barrero, M.~G. Ortega, and C.~Martin, ``Harmonic analysis of direct digital control of voltage inverters,'' \emph{Mathematics and Computers in Simulation}, vol. 130, pp. 155--166, 2016.

\bibitem{gonccalves2019finite}
P.~Gon{\c{c}}alves, S.~Cruz, and A.~Mendes, ``Finite control set model predictive control of six-phase asymmetrical machines - an overview,'' \emph{Energies}, vol.~12, no.~24, p. 4693, 2019.

\bibitem{yao2023weighting}
C.~Yao, G.~Ma, Z.~Sun, J.~Luo, G.~Ren, and S.~Xu, ``Weighting factors optimization for fcs-mpc in pmsm drives using aggregated residual network,'' \emph{IEEE Transactions on Power Electronics}, vol.~39, no.~1, pp. 1292--1307, 2023.

\bibitem{liu2021neural}
X.~Liu, L.~Qiu, W.~Wu, J.~Ma, Y.~Fang, Z.~Peng, and D.~Wang, ``Neural predictor-based low switching frequency fcs-mpc for mmc with online weighting factors tuning,'' \emph{IEEE Transactions on Power Electronics}, vol.~37, no.~4, pp. 4065--4079, 2021.

\end{thebibliography}

\vspace{11pt}

\vfill

\end{document}